\documentclass[12pt,onecolumn,amsfonts]{revtex4-1}
\usepackage{amsfonts}
\usepackage{amsmath}
\usepackage{amssymb}
\usepackage{graphics,graphicx}
\usepackage{epsf}
\usepackage{hyperref}
\usepackage[all]{hypcap}

\newcommand{\be}{\begin{equation}}
\newcommand{\ee}{\end{equation}}
\newcommand{\bea}{\begin{eqnarray}}
\newcommand{\eea}{\end{eqnarray}}
\newcommand{\eqqm}{\end{multline}}

\begin{document}

\author{H. Landa$^1$, M. Drewsen$^2$, B. Reznik$^1$ and A. Retzker$^{3,4}$}
\title{Modes of Oscillation in Radiofrequency Paul Traps}
\affiliation{$^1$School of Physics and Astronomy, Raymond and Beverly Sackler Faculty of Exact Sciences, Tel-Aviv University, Tel-Aviv 69978, Israel.
\\$^2$QUANTOP, Danish National Research Foundation Center for Quantum Optics, Department of Physics and Astronomy,University of Aarhus, DK-8000 \r{A}rhus C., Denmark.
\\$^3$Institut f{\"u}r Theoretische Physik, Universit{\"a}t Ulm, D-89069 Ulm.
\\$^4$Racah Institute of Physics, The Hebrew University of Jerusalem, Jerusalem 91904, Israel}

\begin{abstract}

We examine the time-dependent dynamics of ion crystals in radiofrequency traps. The problem of stable trapping of general three-dimensional crystals is considered and the validity of the pseudopotential approximation is discussed. We derive analytically the micromotion amplitude of the ions, rigorously proving well-known experimental observations. We use a method of infinite determinants to find the modes which diagonalize the linearized time-dependent dynamical problem. This allows obtaining explicitly the (`Floquet-Lyapunov') transformation to coordinates of decoupled linear oscillators. We demonstrate the utility of the method by analyzing the modes of a small `peculiar' crystal in a linear Paul trap. The calculations can be readily generalized to multispecies ion crystals in general multipole traps, and time-dependent quantum wavefunctions of ion oscillations in such traps can be obtained.

\end{abstract}
\maketitle

\section{Introduction}\label{Sec:Intro}


The trapping of charged particles by electromagnetic fields is an essential tool for many investigations within physics \cite{major2005charged,werth2009charged} and chemistry \cite{GerlichChem,Chemical_applications}, as well as for studies of bio-molecules \cite{SchilerProteinsPRL,SchillerProteins}. In the Paul trap, charged particles are confined by oscillating multipole radiofrequency (rf) fields \cite{GerlichChem}. Different types of Paul traps have been proposed and constructed over the years, and an increasing number of experimental and theoretical work is dedicated to the improvement of these devices (see, e.g. \cite{AnomalousHeatingRate}). With the advent of laser-cooling techniques \cite{metcalf1999laser}, a lot of effort is focused on the study of crystalline forms of trapped ions \cite{Walther1987,PhysRevLett.59.2935,PhysRevA.44.4506,Hasse1990419,PhysRevLett.70.818,DubinColdFluidPRL,DubinSchifferModes,DubinCorrelations,PhysRevLett.88.125002,PhysRevLett.88.205003,PhysRevLett.91.165001,SchifferDrewsenHeating,Mitchell13111998,Itano30011998,PhysRevLett.81.2878,PhysRevLett.86.1994,PhysRevLett.91.095002,PhysRevA.66.013412,0953-4075-36-3-310,PhysRevLett.101.260504,Drewsen22Mode,Rosenband28032008,2009NatPh.5.494H,2011NaPho.5.633A,Home04092009,PhysRevA.62.011401,2010NatPh.6.271S,2010NatPh.6.275S,Drewsen_Long_Range_Order}. Single trapped ions have been cooled to the quantum ground state of motion \cite{PhysRevLett.62.403,PhysRevLett.83.4713}, and crystals of a few ions have been cooled to the ground state in at least a few of the motional modes \cite{PhysRevLett.81.1525}. Long-range order has been observed with large structures of thousands of ions in Penning \cite{Mitchell13111998,Itano30011998} and Paul traps \cite{Drewsen_Long_Range_Order,0953-4075-40-15-F01}.

In this contribution we analyze the dynamics of crystals of charged particles confined in rf traps. In Sec.~\ref{Sec:PaulOverview} we review some previous results on the trapping of ions in rf traps. In Sec.~\ref{Sec:Paul} we discuss the linear stability of ion crystals in general, and consider limits of validity of the pseudopotential (time-independent) approximation, in relation to symmetries of the trapped crystal. We derive the micromotion amplitude of ions in a general periodic solution in a Paul trap, and relate our findings to recent experimental results. We then expand the motion of the trapped ions in coordinates of small displacements about the periodic solution, which is the dynamic equivalent of a minimum of the trapping potential. This expansion is completely general and can be applied to any periodic trapping field. In Sec.~\ref{Sec:Solution} we solve the coupled motion of the ions in the time-dependent potential. We find the modes which diagonalize the dynamical problem and obtain explicitly a time-dependent transformation to coordinates in which the motion is that of decoupled linear oscillators. Using this expansion the exact time-dependent wavefunctions of ions in rf traps can be obtained. We present a numerical study of a small crystal in a linear Paul trap in Sec.~\ref{Sec:Numerics}, and conclude in Sec.~\ref{Sec:Conclusion} with comments on further possible applications and research.

\section{Overview of Paul Trapping}\label{Sec:PaulOverview}

The first crystallization of a cloud of charged aluminum microparticles has been reported in \cite{FirstCrystal}. Wigner crystals of 2 to approximately 100 trapped ions in a Paul trap were reported in \cite{Walther1987,PhysRevLett.59.2935} and further investigated in \cite{Walther1988} both experimentally and  numerically. The simulations included the rf trapping, Coulomb interaction, laser cooling and random noise. Depending on trap parameters, the ions have been found to equilibrate either as an apparently chaotic cloud or in an ordered structure. The latter is defined as the `crystal' solution when it is a simple limit cycle, with the ions oscillating at the rf frequency about well-defined average points. The transition between the two phases has been investigated, it was shown that both phases can coexist and hysteresis in the transition has been observed \cite{Walther1988}.

The motion of two ions in the Paul trap has been investigated in detail in various publications  \cite{Brewer1988,Blumel1989,Brewer1990,Baumann1992,BaumannComment1,BaumannComment2,BaumannReply,Frequency_Locked_Orbits1993,Frequency_Locked_Orbits1994,Brewer1994}. In addition to the aforementioned phases, frequency-locked periodic attractors (where the nonlinearity pulls the motional frequencies into integral fractions of the external rf frequency) were found in numerical simulations and experiments. These solutions are different from the crystal in that the ions move in extended (closed) orbits in the trap, whose period is a given multiple of the rf period. However many of these frequency locked solutions are unstable, especially those of a large period, and perturbations such as those coming from the nonlinearity of the laser-cooling mechanism, tend to destroy them. 

Despite the large amplitude motion, the frequency-locked solutions, being periodic, are of course not chaotic. However, even at the presence of cooling, some solutions in the ion trap may behave chaotically for exponentially long times. The authors of \cite{riddled_basins} suggest that, \textit{eventually}, all trajectories settle at frequency locked attractors, at least for two ions at $a=0$. Numerical simulations and experiments with more ions suggest that in general, different types of solutions - of chaotic and of long-range order nature, can coexist at the same parameter values \cite{Walther1988}.

A further important property of two-ion crystals was discovered in \cite{Blumel1993}. The entire first Mathieu stability zone (which corresponds to stable trapping of one ion), was mapped using numerical simulations to determine the stability of the two-ion crystal. It turned out that the Coulomb interaction destabilizes completely the two-ion crystal at some parameter areas. About a year later, it was independently discovered in \cite{Blumel1995,Blumel1995b} and \cite{Brewer1995}, that two ions in a hyperbolic Paul trap may crystallize in a `peculiar' crystal.

The reason that such crystals were termed `peculiar', is that for a two-ion crystal with no angular momentum about the axial $x$-axis, there are only two possibilities in the harmonic pseudopotential approximation (excluding the case of degenerate secular frequencies). The crystal can either align with the $x$-axis or lie in the \textit{y-z} plane. In the latter case we may assume that the crystal is aligned with the $y$-axis, and in both cases the $z$ coordinate can be eliminated. Therefore a crystal which is at angle to both reduced axes $x$ and $y$, i.e. one which forms an angle with both the axial axis and the plane of symmetry, is peculiar.

In \cite{Blumel1995,Blumel1995b}, an improved pseudopotential approximation was derived for two ions in the hyperbolic Paul trap (and in \cite{Drewsen2000} for the linear Paul trap). This nonlinear pseudopotential can be used to derive approximately the areas of stability in parameter space of the axial and radial two-ion crystal, and the orientation of the peculiar crystal. However, even the improved nonlinear pseudopotential cannot reproduce the areas of no stable crystal.

As mentioned above, the crystal solution in all of these works, is taken to be a periodic solution of the equations of motion with a period equal to that of the rf potential. This is a limit cycle of the equations, the simplest of the frequency locked solutions mentioned above, and it is an attractive one when cooling is present and it is stable. Its stability is analyzed in \cite{Brewer1995} by looking at the Poincar\'e map associated with this solution. The Poincar\'e map is a mapping of the phase space into itself, in which every inital point $\left\{x\left(0\right),p\left(0\right)\right\}$ is evolved according to the equations of motion and mapped to $\left\{x\left(T\right),p\left(T\right)\right\}$ after one rf period $T$. The crystal is by definition a fixed point of the Poincar\'e map, and its linear stability is determined by the linearized dynamics in its vicinity, specifically by the eigenvalues of the so-called monodromy matrix. The linearization takes into account the leading (harmonic) coupling between the ions, expanded about the periodic solution. Further analysis (both linear and nonlinear) of the periodic orbit of two ions in a Paul trap appears in \cite{Vogt1994,Vogt1994b,Blumel_rf_resonances,periodic_motion_of_two_ions}.

\section{Paul Traps}\label{Sec:Paul}
 

\subsection{The Trapping Potential}\label{Sec:PaulIntroduction}

We start with the potential energy of identical ions in the most general single, nonsegmented quadrupole rf trap. By choosing a specific frequency $\bar{\omega}$, which is characteristic of the secular frequencies in the problem, and measuring time and distances in units of $1/\bar{\omega}$ and $d=\left(e^{2} /m\bar{\omega}^{2} \right)^{1/3}$, respectively (with $m$ and $e$ the ion's mass and charge), we write the nondimensional potential as
\be {V} = {V}_{\rm trap} + {V}_{\rm Coulomb} =\sum _{i}^{N}\frac{1}{2} \left(\Lambda_{x} x_{i} ^{2} +\Lambda_{y} y_{i} ^{2} +\Lambda_{z} z_{i} ^{2} \right) + \sum _{i\ne j}\frac{1}{2} \left\| \vec{R}_{i} -\vec{R}_{j} \right\| ^{-1}, \label{Eq:VNondimful}\ee
where $\vec{R}_{i} = \left\{x_{i},y_{i},z_{i}\right\}$ is the vector coordinate of ion $i$, the trapping terms are given by ${\Lambda }_{\alpha } =\frac{\Omega^2}{4}\left(a_{\alpha } -2q_{\alpha } \cos\Omega t\right)$, $\alpha \in \left\{x,y,z\right\}$, with $a_{\alpha }$ and $q_{\alpha }$ being the nondimensional Mathieu parameters of the respective coordinates \cite{WinelandReview}, and ${\Omega}$ being the rf frequency (in units of $\bar{\omega}$).

Regarding the trapping parameters, the Laplace equation implies that
\be \sum _{\alpha }a_{\alpha }  = \sum _{\alpha }q_{\alpha } =0 \label{Eq:Laplace}.\ee
Two commonly used types of traps are the hyperbolic \cite{Ghosh} and the linear \cite{WinelandReview} Paul traps. Taking the $x$-axis as the axial direction, the hyperbolic trap can be described by setting $a_{y} =a_{z} \equiv a$, $a_{x} =-2a$ and $q_{y} =q_{z} \equiv q$, $q_{x} =-2q$, while in the linear trap we have $a_{y} =a_{z} \equiv a$, $a_{x} =-2a$ and $q_{y} =-q_{z} \equiv q$, $q_{x} =0$ (so $a$ must be negative to obtain stable trapping). In the next subsection we will relate to symmetries of these two trapping geometries, but for now we keep the discussion completely general.

The trapping potential ${V}_{\rm trap}$ when considered alone, gives rise to decoupled Mathieu equations in each ion coordinate \cite{McLachlan}.
The corresponding characteristic exponents are $\beta _{\alpha} \left(a_{\alpha},q_{\alpha}\right)$ and the secular frequencies are then ${\omega}_{\alpha} \equiv \beta_{\alpha} \frac{{\Omega}}{2}$. In these units, ${V}_{\rm trap}$ and ${V}_{\rm Coulomb}$ are both of order unity for a crystal, when the trapping balances the Coulomb repulsion. These units allow naturally the introduction of the parameter 
\be \epsilon \equiv 4/\Omega^2 \label{Eq:epsilon}.\ee
If $\bar{\omega}$ is the (dimensional) secular frequency along some specific axis $\alpha$, we have ${\omega}_{\alpha}=1$ and $\epsilon=\beta_{\alpha}^2$. Using the familiar lowest order approximation \cite{WinelandReview}, $\beta _{\alpha} \approx \sqrt{a_{\alpha}+q_{\alpha}^{2} /2} $, we see that the limit $\epsilon \to 0$ is equivalent $a \to 0$, $q^2 \to 0$. This parameter will be used in the following.

When the Mathieu parameters belong to a single-particle stability-zone, the potential of eq.~\eqref{Eq:VNondimful} allows for bounded motion and large ion clouds can be trapped for extraordinary long times without cooling. This is given by the limit $\vec{R}_i \to \infty$ in eq.~\eqref{Eq:VNondimful}, which means that ions are not crystallized, rather their motion is that of decoupled trapped particles. Despite the large amplitude motion, at low enough density the ions hardly interact.

However, stable single-particle parameters do not guarantee that a stable crystal solution exists, even with two ions, as discussed extensively in Sec.~\ref{Sec:PaulOverview}, which has also been recently investigated with crystals consisting of upto several hundreds of ions \cite{DrewsenRingLikeStability,DrewsenDisksPRL}. We recall that a stable crystal solution is considered as a periodic solution with the same period of the rf, which is linearly stable under small perturbations. The existence of a stable crystal solution is a property of the fully nonlinear and time-dependent problem. There may also exist stable multi-periodic solutions (with a period which is some large, arbitrary multiple of the rf period), the frequency-locked solutions described in Sec.~\ref{Sec:PaulOverview}.

In experiments, even at the presence of cooling, crystals may exist only metastably, changing completely after a given time or when parts of the crystal are moving with respect to the rest \cite{Drewsen_Long_Range_Order}. In this case, if the timescale of the change of the crystal is long enough, the analysis in the following subsections, using modes expanded about the (quasi-) periodic solution may still be useful, even though at least one mode would be unstable.

\subsection{Linearization using the Pseudopotential Modes}\label{Sec:PseudoLin}

In this subsection we linearize the crystal motion using the pseudopotential modes. Therefore this treatment assumes that a crystal solution exists, and is close to the crystal of the pseudopotential. As known from the example of two-ions in a hyperbolic Paul trap, this assumption may not hold, even for arbitrarily small values of $a$ and $q$ (where the crystal is `peculiar' \cite{Blumel1995,Blumel1995b,Brewer1995}). The linearized secular modes of a one-dimensional Coulomb chain were treated in detail in \cite{Peeters1D,MorigiFishmanModesPRL,MorigiFishmanModesPRE}.

Let us define the squared ratio of secular radial and axial frequencies $\gamma_y=\omega_y^2/\omega_x^2$, $\gamma_z=\omega_z^2/\omega_x^2$ and $\gamma_x=\omega_x^2/\omega_x^2=1$, thus choosing $\bar{\omega}$ as the axial trapping frequency, so we can rewrite the potential as the sum ${V} = {V}_{\rm pseudo} + {V}_{\rm Coulomb} + {V}_{\rm rf} $,
\be {V} = \sum _{i,\alpha}^{N}\frac{1}{2} \gamma_{\alpha} \left(\vec{R}_{i,\alpha} \right)^{2} + \sum _{i\ne j}\frac{1}{2} \left\| \vec{R}_{i} -\vec{R}_{j} \right\| ^{-1} +\sum _{i,\alpha}^{N}\frac{1}{2} \left(\Lambda_{\alpha}-\gamma_{\alpha}\right) \left(\vec{R}_{i,\alpha}\right) ^{2}.\label{Eq:Vnondimful2} \ee

We expand about the minimum-configuration locations, $\left\{\vec{R}_{i} ^{0} \right\}$, obtained from the secular part of $V$ in eq.~\eqref{Eq:Vnondimful2}, ${V}_{\rm pseudo} + {V}_{\rm Coulomb}$, and change to the normal modes $\Theta _{j} $ by setting 
\be\vec{R}_{i,\alpha} \left(t\right)=\vec{R}_{i,\alpha} ^{0} +\sum _{j}^{3N}D_{i,\alpha}^{j} \Theta _{j} \left(t\right),\ee
where $D_{i,\alpha}^{j} $ is the matrix of normal mode vectors, with rows indexed by the $N$ ion numbers $i$ and $3$ directions $\alpha$, and columns by the $3N$ normal mode numbers $j$. Writing the potential in terms of the normal modes, $V={V}_{\rm harmonic} +{V}_{\rm rf}+ V_{\rm nonlinear}$, we keep only the first two terms to get
\be V=\sum _{i}\frac{1}{2} \omega _{i} ^{2} \Theta _{i} ^{2} +\sum _{i,\alpha}^{N}\frac{1}{2} \left(\Lambda_{\alpha} -\gamma_{\alpha}\right)\left(\vec{R}_{i,\alpha} ^{0} +\sum _{j}^{3N}D_{i,\alpha}^{j} \Theta _{j}  \right)^{2}+... \label{Eq:VrfModes} \ee
and the linearized equation of motion (e.o.m) derived from eq.~\eqref{Eq:VrfModes} is
\be \ddot{\Theta}_m+{\omega}_m^2{\Theta}_m=-\frac{{\partial V}_{\rm rf}}{\partial\Theta_m} =-\sum _{i,\alpha}^{N}\left(\Lambda_{\alpha} -\gamma_{\alpha}\right)D_{i,\alpha}^{m} \left(\vec{R}_{i,\alpha} ^{0} +\sum _{j}^{3N}D_{i,\alpha}^{j} \Theta _{j}  \right). \label{Eq:rfeomModes} \ee

Before treating the coupled system described by eq.~\eqref{Eq:rfeomModes}, we first note that when the rf trapping potential is symmetric with respect to the axes of symmetry of the crystal, the equations of motion are in fact diagonal. For the following few paragraphs, let us therefore limit the discussion to the hyperbolic and the linear Paul traps, whose parameters were described following eq.~\eqref{Eq:Laplace}. Given these two trapping geometries, the current expansion will be diagonal if the crystal is two-dimensional and lies in the plane of cylindrical symmtery of a hyperbolic trap, or the crystal is one-dimensional and aligned with one of the trap axes, in either the hyperbolic or the linear trap. In that case, we have $\vec{R}_{i,\alpha} ^{0} =0$ for the coordinates transverse to the crystal, and also the normal modes decouple in these directions from the modes tangential to the crystal (i.e. $D_{i,\alpha}^{j}$ is divided into blocks in the index $\alpha$).

We then can use two relations which hold in the plane of the crystal or along its axis,
\be \sum _{i,\tilde{\alpha}}D_{i,\tilde{\alpha}}^{m} D_{i,\tilde{\alpha}}^{j} = \delta_{mj}, \qquad \qquad \sum_{i,\tilde{\alpha}}D_{i,\tilde{\alpha}}^{m}\vec{R}_{i,\tilde{\alpha}}^{0}=\xi_b\delta_{mb},\label{Eq:Relations}\ee
where hereafter, $\tilde{\alpha}$ runs on the symmetry directions , $\xi_b= \sqrt{\sum_{i,\tilde{\alpha}}\left(\vec{R}_{i,\tilde{\alpha}}^{0}\right)^2}$ and $b$ denotes the breathing mode. The identity on the left in eq.~\eqref{Eq:Relations} is the completeness of the normal modes along the symmetry directions. To get the second identity we use the fact that in a harmonic trap, the breathing mode vector is exactly proportional to the minimum-configuration locations $\vec{R}_{i,\alpha}^{0}$ of the pseudopotential (see, e.g. App. B of \cite{DubinSchifferModes}), so that $D_{i,\tilde{\alpha}}^{b} = \vec{R}_{i,\tilde{\alpha}}^{0}/\xi_b$, and therefore the vector $\vec{R}_{i,\tilde{\alpha}}^{0}$ is orthogonal to all other normal mode vectors. By using eq.~\eqref{Eq:Relations} we can replace the r.h.s in eq.~\eqref{Eq:rfeomModes} with the simple expression
\be -\frac{\partial V_{\rm rf}}{\partial\Theta_m} =-\left(\Lambda_{\tilde{\alpha}} -\gamma_{\tilde{\alpha}}\right) \left(\xi_b\delta_{mb} + \Theta _{m}  \right).\ee

We find that the equations of motion for modes in the crystal plane or along its axis are, after multiplying by $\epsilon=4/\Omega^2$ of eq.~\eqref{Eq:epsilon} and rescaling $t \to \Omega t/2$,
\be \ddot{\Theta}_m + \left[\epsilon\left(\omega_m^2 - \gamma\right) +\left(a-2q\cos2t\right)\right]\Theta_m = \left[\epsilon\gamma -\left(a -2q\cos2t\right)\right]\xi_b\delta_{mb} \label{Eq:eomModes}\ee
where $\gamma$, $a$ and $q$ are defined along the symmetry axes. Similar equations hold in the directions transverse to the crystal, without the inhomogenous r.h.s and with the corresponding $\gamma$, $a$ and $q$.

Eq.~\eqref{Eq:eomModes} shows explicitly how the isotropy of the potential along the axes of symmtery of the crystal allows to decouple the equations of the modes. This decoupling puts the conditions for linear stability of the crystal in terms of diagonal Mathieu equations for each of the modes. In addition, with this symmetry the only mode with an inhomogenous r.h.s is the breathing mode, i.e. it is the only mode on which the rf potential acts as a driving force. This driven motion is in fact the $\pi$-periodic motion of the crystal as whole, about the static minimum-configuration locations of the pseudopotential, $\left\{\vec{R}_{i} ^{0} \right\}$.

Returning to the case of a general trap, in order to put eq.~\eqref{Eq:rfeomModes} in a simple form we multiply it by $\epsilon=4/\Omega^2$ of eq.~\eqref{Eq:epsilon} and rescale $t \to \Omega t/2$. Defining the two matrices
\be A_{mj} =\epsilon\omega_m^2\delta _{mj} + \sum _{i,{\alpha}}D_{i,{\alpha}}^{m} D_{i,{\alpha}}^{j}  \left(a_{\alpha }-\epsilon\gamma_{\alpha } \right), \qquad
Q_{mj} =\sum _{i,{\alpha}}D_{i,{\alpha}}^{m} D_{i,{\alpha}}^{j}  q_{\alpha } \label{Eq:AQmodes}\ee 
and the vectors
\be G_{m}  = -\sum _{i,{\alpha}}D_{i,{\alpha}}^{m} \left(a_{\alpha }-\epsilon\gamma_{\alpha } \right) \vec{R}^{0} _{i,\alpha }, \qquad F_{m} = \sum _{i,{\alpha}}D_{i,{\alpha}}^{m} q_{\alpha } \vec{R}^{0} _{i,\alpha }, \label{Eq:GFmodes}\ee 
we rewrite eq. \eqref{Eq:rfeomModes} in vector notation as
\be\ddot{\vec{\Theta}} +\left[A -2Q\cos 2t\right]\vec{\Theta} = \vec{G} +2\vec{F}\cos2t.\label{Eq:eomModesInHomogenous} \ee 

We solve the homogeneous l.h.s of eq.~\eqref{Eq:eomModesInHomogenous} in Sec.~\ref{Sec:Solution}. Since the pseudopotential modes are expanded aboout static configuration points, eq.~\eqref{Eq:eomModesInHomogenous} is an inhomogenous equation with a $\pi$-periodic r.h.s, which has a unique $\pi$-periodic solution (except for possibly a region of measure-zero in $a_{\alpha},q_{\alpha}$ space). This periodic solution of driven motion of the normal modes, corresponds to the exact $\pi$-periodic solution which defines the crystal in the rf trapping potential. Details of the solution of the inhomogeneous equation can be found in \cite{rfmodes}.

Eq.~\eqref{Eq:eomModesInHomogenous} shows that in general, the rf couples the pseudopotential normal modes and also acts as a driving force. Under this coupling, the true modes of oscillation of the system may in general have different frequencies (and even lose stability), and different oscillation directions than the pseudopotential modes upon which this expansion is based. Indeed, the linearization starting from the pseudopotential approximation may not be adequate in the general case. We investigate this point further in the following two subsections.

\subsection{The Periodic Crystal Solution}\label{Sec:Periodic}

We now abandon the pseudopotential approximation and turn to studying the time-dependent potential directly. In this subsection we derive analytically the micromotion amplitude of the ions in typical crystals in Paul traps. The e.o.m for the ion coordinates, derived from eq.~\eqref{Eq:VNondimful} after rescaling by $t \to \Omega t/2$, is
\be\ddot{\vec{R} }_{i,\alpha } + \left( a_{\alpha } -2q_{\alpha } \cos2 t\right) \vec{R} _{i,\alpha } -\epsilon\sum _{\substack{ j=1\\j\ne i}}^N \left\|\vec{R}_{i} -\vec{R}_{j}\right\|^{-3}\left(\vec{R}_{i,\alpha } -\vec{R}_{j,\alpha}\right)=0. \label{Eq:eomZeta1}\ee
Eq.~\eqref{Eq:eomZeta1} has $\pi$-periodic coefficients and is time-reversal invariant. We assume the existence of a crystal in the sense of Sec.~\ref{Sec:PaulOverview}, i.e. a $\pi$-periodic and time-reversal invariant solution, which obtains the genreal form
\be \vec{R}_{i,\alpha}^{\pi}\left(t\right) = \sum_{n=-\infty}^{n=\infty}\vec{B}_{2n,i,\alpha} e^{i2nt} \label{Eq:Rpi}.\ee
In this form, the average ion location is $\vec{B}_{0,i}$ (which is different from $\vec{R}_{i}^{\pi}\left(t=0\right)$).

We wish now to see what can be said about $\vec{R}_{i}^{\pi}$ in typical Paul trapping experiments. We first define the dynamic matrix 
\be G_{ij}\left(\left\{\vec{R}_{i}\left(t\right)\right\}\right)= \delta_{ij}\sum _{\substack{m\\m\ne i}} \left\|\vec{R}_{i} -\vec{R}_{m}\right\|^{-3} -\left(1- \delta_{ij}\right)\left\|\vec{R}_{i} -\vec{R}_{j}\right\|^{-3}, \label{Eq:Gij}\ee
and write using eq.~\eqref{Eq:Rpi} its Taylor and Fourier expansion around $\left\{\vec{B}_{0,i}-\vec{B}_{0,j}\right\}$ as
\be G_{ij}= G_{0,ij}\left(\left\{\vec{B}_{0,i}\right\}\right)+G_{2,ij}\left(\left\{\vec{B}_{0,i}\right\},\left\{\vec{B}_{2,i}\right\}\right)\left(e^{2it}+e^{-2it}\right)+... \label{Eq:GijExpansion}\ee

Substituting the solution ansatz eq.~\eqref{Eq:Rpi} into eq.~\eqref{Eq:eomZeta1} we get
\be\sum_n\left[ \left( a_{\alpha } -\left(2n\right)^2\right)\vec{B} _{2n,i,\alpha } -q_{\alpha } \left(\vec{B} _{2n-2,i,\alpha }+\vec{B} _{2n+2,i,\alpha }\right) -\epsilon\sum _{j} G_{ij}\vec{B} _{2n,j,\alpha }\right]e^{i2nt} =0. \label{Eq:eomZeta111}\ee
We replace $G_{ij}\left(\left\{\vec{R}_{i}\left(t\right)\right\}\right)$ in eq.~\eqref{Eq:eomZeta111} by its leading order, time-independent term from eq.~\eqref{Eq:GijExpansion}, $G_{0,ij}=G_{ij}\left(\left\{\vec{B}_{0,i}\right\}\right)$, and require that the above relation holds for every $t$. We get, by using $\vec{B} _{2n}=\vec{B}_{-2n}$ and neglecting $\vec{B}_{4n}\approx 0$ (which is of the next order in $q_{\alpha}$), the equation coming from the $n=0$ term,
\be a_{\alpha }\vec{B}_{0,i,\alpha}-2q_{\alpha }\vec{B}_{2,i,\alpha}-\epsilon \sum_j G_{0,ij}\vec{B}_{0,j,\alpha} = 0,\label{Eq:B0B2n0}\ee
and the coupled equation from the coefficient of $e^{2it}$,
\be \left(a_{\alpha }-4\right)\vec{B}_{2,i,\alpha}-q_{\alpha }\vec{B}_{0,i,\alpha}-\epsilon \sum_j G_{0,ij}\vec{B}_{2,j,\alpha} = 0 \label{Eq:B0B2n2}.\ee

This sytem of equations can be seen as a linear homogenous system in the $2N$ variables $\left\{\vec{B}_{0,i,\alpha},\vec{B}_{2,i,\alpha}\right\}$, $i=1,...,N$, for fixed $\alpha$, since the equations are diagonal along the three different axes. Of course, this is not really a linear system since the matrix $G_{0,ij}$ depends on $\vec{B}_{0,i}$, but since we are not trying to actually solve the system, this will not matter. Let us fix the index $\alpha$ to some axis (supressing it in the following), and define the $N$-component vectors $\vec{u}_{0,i}=\vec{B}_{0,i,\alpha}$ and $\vec{u}_{2,i}=\vec{B}_{2,i,\alpha}$. Then the above system can be written in block-matrix form
\be\left(\begin{array}{cc} {a-\epsilon G_0} & {-2q } \\ {-q} & {-4} \end{array}\right)\left(\begin{array}{c} {\vec{u}_0} \\ {\vec{u}_2} \end{array}\right)=0, \label{Eq:B0B2system}\ee
where we have neglected $a-\epsilon G_0$ as compared with $-4$ in the lower-right block of eq.~\eqref{Eq:B0B2system}. Since we assume that the system has a solution (which approximates the $\pi$-periodic crystal), the above matrix must be singular. We can expand its determinant,
\be 0=\det\left[a-\epsilon G_0 -\left(-2q\right)\left(-4\right)^{-1}\left(-q\right)\right]=\det\left[a-\epsilon G_0 +q^2/2\right].\label{Eq:B0B2det}\ee
Taking $\vec{u}_0$ to be the vector from the kernel of $\left(a-\epsilon G_0 +q^2/2\right)$ which obeys $\left(a-\epsilon G_0\right)\vec{u}_0=-\left(q^2/2\right)\vec{u}_0$, we find that the solution of eq.~\eqref{Eq:B0B2system} is $\vec{u}_2=-\left(q/4\right)\vec{u}_0$.

We therefore have obtained that in a general quadrupole trap, in the $\pi$-periodic crystal solution of eq.~\eqref{Eq:Rpi}, every ion coordinate obeys (at least to leading order in $a_{\alpha}/4,q_{\alpha}/4,\epsilon_{\alpha}/4$),
\be\vec{B}_{2,i,\alpha}\approx-\frac{q_{\alpha}}{4}\vec{B}_{0,i,\alpha},\label{Eq:B2qB0}\ee
i.e. that the micromotion amplitude in each coordinate is $q_{\alpha}/2$ of the respective average position, and at $\pi$ phase with respect to the rf drive. In the hyperbolic trap the corresponding motion has been imaged as early as in \cite{FirstCrystal}. In simulations of a generic trap (with different $q$ and $a$ parameters for the three axes), eq.~\eqref{Eq:B2qB0} seems to hold accurately to within a few percent, for $q$ up to $\sim 0.7$, which is consistent with a deviation of order $\left(q_{\alpha}/4\right)^2$.

The relation in eq.~\eqref{Eq:B2qB0} loses its accuracy when either $\vec{B}_{0,i,\alpha}\ll 1$ or $q_{\alpha}\ll 1$. In the former case, for an ion near the zero of one of the rf axes, the corresponding micromotion amplitude in fact seems in the cases we have checked, to be lower (this is similar to a single trapped ion at the origin of a Paul trap, for which the unique $\pi$-periodic solution is the trivial one).

For the linear Paul trap case of $q_{\alpha}\ll 1$ along the axial axis, the first-order expression in eq.~\eqref{Eq:B2qB0} loses its meaning. We must therefore add to eq.~\eqref{Eq:B0B2n2} the second term in eq.~\eqref{Eq:GijExpansion}, $G_{2,ij}$, which rotates at the frequency $e^{2it}$. Setting $q=0$, eq.~\eqref{Eq:B0B2system} is replaced with 
\be\left(\begin{array}{cc} {a-\epsilon G_0} & {0 } \\ {-\epsilon G_{2}} & {-4} \end{array}\right)\left(\begin{array}{c} {\vec{u}_0} \\ {\vec{u}_2} \end{array}\right)=0.\label{Eq:B0B2systemPaul}\ee
We now examine what can be deduced about $G_2$. The off-diagonal elements are in general
\be G_{2,ij}= 3 \left\|\vec{B}_{0,i} -\vec{B}_{0,j}\right\|^{-5}\sum_{\alpha}\left(\vec{B}_{0,i,\alpha} -\vec{B}_{0,j,\alpha}\right)\left(\vec{B}_{2,i,\alpha} -\vec{B}_{2,j,\alpha}\right), \quad i\ne j. \label{Eq:G2ij}\ee
In particular, using eq.~\eqref{Eq:B2qB0} with $q_y=-q_z=q$ and $q_x=0$, we get that for the linear Paul trap
\be G_{2,ij}= -3\frac{q}{4} \left\|\vec{B}_{0,i} -\vec{B}_{0,j}\right\|^{-5}\left[\left(\vec{B}_{0,i,y} -\vec{B}_{0,j,y}\right)^2-\left(\vec{B}_{0,i,z} -\vec{B}_{0,j,z}\right)^2\right]+{\rm O}\left(\frac{q^2}{4^2}\right), \quad i\ne j. \label{Eq:G2ijPaul}\ee
Since the diagonal elements are the negative of the row sum, as in eq.~\eqref{Eq:Gij}, $G_{2,ii}=-\sum_{m\ne i}G_{2,im}$, we immediately find that for a crystal which is invariant (up to a permutation of the ions) under $y\leftrightarrow \pm z$, $G_{2,ii}=0+{\rm O}\left({q^2}/{4^2}\right)$. In fact, for the linear Paul trap, the e.o.m, eq.~\eqref{Eq:eomZeta1} is invariant under $y\leftrightarrow -z,t\to t+\pi/2$, so given one crystal solution, there is also a solution related by this transformation. Depending on the number of ions and trapping parameters, both solutions may actually be the same crystal. As the crystal size grows, by the same symmetry arguments, $G_{2,ii}\to 0+{\rm O}\left({q^2}/{4^2}\right)$.

The second row of eq.~\eqref{Eq:B0B2systemPaul} gives $\vec{B}_{2,i,x}=-{\epsilon}/{4}\sum G_{2,ij}\vec{B}_{0,j,x}$, and we argue that by the above symmtery arguments, for a typical symmetric or large crystal in a linear Paul trap, when the crystal configuration is also symmetric under $x\leftrightarrow -x$, the first order terms in the above summation will cancel, and (using $\epsilon\approx q/2$)
\be\vec{B}_{2,i,x}={\rm O}\left(\frac{1}{2}\cdot\frac{q^3}{4^3}\right)\vec{B}_{0,i,x}.\label{Eq:B2q2B0}\ee
If the symmetry alluded to above does not hold, eq.~\eqref{Eq:B2q2B0} should be replaced with an expression which is one order less, namely $\vec{B}_{2,i,x}={\rm O}\left(\frac{\epsilon}{4}\cdot\frac{q}{4}\right)\vec{B}_{0,i,x}$.

Eq.~\eqref{Eq:B2q2B0} explains why in the linear Paul trap, there is essentially no micromotion excitation along the axial direction despite the strong Coulomb interaction. In addition, since $q_y=-q_z$ in the linear trap, the oscillation described by eq.~\eqref{Eq:B2qB0} is exactly the (2,2) mode of cold-fluid theory \cite{DubinColdFluidPRL} which has been observed both in simulations \cite{SchifferDrewsenHeating} and recently discussed in connection with experiments \cite{Drewsen22Mode}. In fact, in simulations we have performed (Sec.~\ref{Sec:Numerics}), eq.~\eqref{Eq:B2qB0} seems to hold extremely accurately radially (to within half a percent), and eq.~\eqref{Eq:B2q2B0} gives an accurate estimate for the axial micromotion amplitude (also matching e.g., \cite{SchifferDrewsenHeating,HeatingSchiller}).

\subsection{Linearization about the Periodic Crystal, and the Pseudopotential Limit}\label{Sec:PointsLin}

In this subsection we expand the modes of oscillation of the ions about the periodic crystal solution, eq.~\eqref{Eq:Rpi}. The following expansion is applicable to general crystal configurations, does not rely on the pseudopotential modes or on a small parameter, and can be readily generalized to settings not considered in this paper, e.g. segmented traps and higher-order multipole traps. We will discuss its relation to the expansion of Sec.~\ref{Sec:PseudoLin}, which used the pseudopotential modes.

Expanding the potential of eq.~\eqref{Eq:VNondimful} about the time-dependent solution $\left\{\vec{R}_{i} ^{\pi}\left(t\right) \right\}$, in the coordinates of small oscillations $r _{i,\alpha } =\vec{R}_{i,\alpha } -\vec{R}_{i,\alpha} ^{\pi}\left(t\right) $, and defining the ${\pi}$-periodic matrix
\be {K}_{ij} ^{\sigma \tau } \left(t\right)=\left. \frac{\partial ^{2} {V}_{\rm Coulomb} }{\partial R_{i,\sigma} \partial R_{j,\tau } } \right|_{\left\{\vec{R}_{i} ^{\pi}\left(t\right) \right\}}, \ee
 we find the linearized e.o.m
\be\ddot{r }_{i,\sigma } + \left( a_{\sigma } -2q_{\sigma } \cos2 t\right) r _{i,\sigma } +\epsilon\sum _{j,\tau } {K}_{ij} ^{\sigma \tau }\left(t\right) r _{j,\tau } = 0. \label{Eq:eomZeta2}\ee 

This equation is a linearly coupled system for the ion coordinates with $\pi$-periodic coefficients. We see that the interaction term in eq.~\eqref{Eq:eomZeta2} is of the same order (in $\epsilon$, or equivalently $a$ and $q^2$) as the diagonal term. Expanding the matrix ${K}_{ij} ^{\sigma \tau }\left(t\right)$ in a Fourier series in the form 
\be{K}_{ij} ^{\sigma \tau }=\left({K}_0\right)_{ij} ^{\sigma \tau }-2\left({K}_2\right)_{ij} ^{\sigma \tau }\cos 2t-...,\label{Eq:KijFourier}\ee
we first define the two matrices
\be A_{ij} ^{\sigma \tau } =\delta _{ij} \delta _{\sigma \tau } a_{\sigma } + \epsilon \left({K}_0\right)_{ij} ^{\sigma \tau }, \qquad
Q_{ij} ^{\sigma \tau } =\delta _{ij} \delta _{\sigma \tau } q_{\sigma } + \epsilon \left({K}_2\right)_{ij} ^{\sigma \tau }.\label{Eq:AQ}\ee 
We can now switch notation to a simpler one with dynamical variables $u_{m} $ whose single index $m$ stands for both indices $\left\{i,\sigma \right\}$ of $r _{i,\sigma } $, with the corresponding indices replaced in eq.~\eqref{Eq:AQ}, and rewrite eq. \eqref{Eq:eomZeta2} in vector notation, with only the two leading harmonics of the Fourier series, as
\be\ddot{\vec{u}} +\left[A -2Q\cos 2t\right]\vec{u} = 0.\label{Eq:eomuInHomogenous} \ee 
This equation describes linealrized coupled perturbations about the $\pi$-periodic solution which defines the crystal in the time-dependent potential. In Sec.~\ref{Sec:Solution} we will solve this coupled system and find its decoupled modes of oscillation.

Eq.~\eqref{Eq:eomuInHomogenous} is seen to be identical in form to the l.h.s of eq.~\eqref{Eq:eomModesInHomogenous}, which is given in terms of the pseudopotential modes. However, the current expansion is based on the exact force acting between the ions during their motion along the periodic trajectory of the full potential. Higher harmonics from the Fourier series of this force can be added, and in Sec.~\ref{Sec:Numerics} we will in fact include the next harmonic ($\cos4t$), to obtain very accurate results for the modes. When the nonlinear motion is not exactly $\pi$-periodic, some of the linearized modes of the above expansion may not be oscillatory (as detailed in Sec.~\ref{Sec:Floquet}), which describes (locally) the aperiodic motion of the crystal. If this change of the crystal is slow, the linearization of eq.~\eqref{Eq:eomZeta2} can still be useful. 

Considering the limit of taking the Mathieu parameters to zero, this does not guarantee that the crystal will be (linearly) stable. If there are unstable modes, they may remain unstable even as $a,q^2,\epsilon\to 0$. As for approaching the pseudopotential crystal in this limit, \textit{if} the average ion locations $\left\{\vec{B}_{0,i} \right\}$ of the periodic solution tend to the pseudopotential minimum locations, then the limit of the pseudopotential modes is regained on the l.h.s of eq.~\eqref{Eq:eomZeta2}. Characterizing the conditions for this to occur requires further investigations, however for simple cases the results of Sec.~\ref{Sec:Periodic} may provide some hints.

In particular, if the rf potential is isotropic with respect to the configuration (as discussed in Sec.~\ref{Sec:PseudoLin} for specific cases), and we neglect $K_2$ in eq.~\eqref{Eq:AQ}, the matrix $Q$ becomes proportional to the identity. Then we can diagonalize eq.~\eqref{Eq:eomuInHomogenous} by an orthogonal transformation (which is exactly the transformation to the normal modes) and obtain eq.~\eqref{Eq:eomModes}. There will now not be an inhomogeneous r.h.s, since the periodic crystal motion is already accounted for. At the presence of driven breathing oscillations, the ions feel stronger repulsive nonlinearity when they oscillate radially inwards, so we may expect the real crystal to be a bit more expanded (with the ion distances somewhat larger) as compared with the pseudopotential approximation.

We note that for an axial chain of ions in the linear Paul trap, indeed $K_2\approx 0$ even if there is a small rf leak in axial direction, or when the ions do not lie exactly along the rf null axis, and eq.~\eqref{Eq:eomModes} is almost exact. For the axial modes, the correction to the pseudopotential modes will be small. For the radial modes, the correction will typically be small, provided that the rf frequency is large compared with the radial frequencies. However, for calculation of small effects which might be of importance in high-accuracy experiments such as for quantum information processing \cite{WinelandReview}, the full time-dependent solution must be used.

As a final note, a generalization of the isotropy of the rf potential with respect to the crystal configuration, would be the existence of a constant orthogonal transformation which diagonalizes eq.~\eqref{Eq:eomuInHomogenous}, and thus decouples the modes in the non-isotropic case. Such a transformation would simultaneously diagonalize the matrices $A$ and $Q$ (which must commute), and does not exist in the general case.

\section{Solution of the Linear Equations}\label{Sec:Solution}

\subsection{The Floquet Problem}\label{Sec:Floquet}

Eq.~\eqref{Eq:eomuInHomogenous} is a linear differential equation with periodic coefficients and therefore amenable to treatment using Floquet theory, which we briefly review here. For more details see \cite{rfmodes} and references within. For the Newtonian problem with $f$ degrees of freedom ($f=3N$ for $N$ ions in three-dimensions), the corresponding Floquet problem is stated in terms of coordinates in $2f$-dimensional phase space by the definitions

\be\phi =\left(\begin{array}{c} {\vec{u}} \\ {\dot{\vec{u}}} \end{array}\right), \quad \Pi \left(t\right)=\left(\begin{array}{cc} {0} & {1_{f} } \\ {-\left(A-2Q\cos 2t\right)} & {0} \end{array}\right), \label{Eq:phi}\ee
where $1_{f} $ is the $f$-dimensional identity matrix. The e.o.m is written in standard form as
\be \dot{\phi }=\Pi \left(t\right)\phi \label{Eq:eomFloquet}. \ee 

In the following, an $f$-dimensional vector $\vec{u}$ will be denoted by a lower case Latin letter with an arrow. $f$-dimensional matrices will be denoted by capital Latin letters ($Q$). A $2f$-dimensional vector $\phi$ will be denoted by a lower case Greek letter. Capital Greek letters (unitalicized) will denote $2f$-dimensional matrices ($\Pi$, {\rm B}).

A fundamental matrix solution to eq.~\eqref{Eq:eomFloquet} has $2f$ linearly independent column solutions and obeys the matrix equation $\dot{\Phi }\left(t\right)=\Pi \left(t\right)\Phi \left(t\right)$. A fundamental matrix solution which equals the identity matrix at $t=0$, i.e. obeys $\Phi \left(0\right)=1_{2f}$, is known as the \textit{matrizant} of eq.~\eqref{Eq:eomFloquet} (and is obviously unique). The matrizant can always be written in the form $\Phi \left(t\right) =\Gamma \left(t\right)e^{{\rm B} t} \Gamma ^{-1}\left(0\right)$ where $\Gamma\left(t+T\right)=\Gamma\left(t\right)$ is periodic with the period $T$ of $\Pi\left(t\right)$, and the constant matrix ${\rm B}$ is diagonal, with entries known as the characteristic exponents of the Floquet problem,
\be{\rm B} ={\rm diag}\left\{i\beta _{1} ,...,i\beta _{2f} \right\}. \label{Eq:B}\ee
The time-dependent linear coordinate change, known as the `Floquet-Lyapunov' transformation, defined by
\be\phi\left(t\right)=\Gamma \left(t\right)\chi\left(t\right), \label{Eq:FLTrans}\ee
transforms the e.o.m eq.~\eqref{Eq:eomFloquet} into the time-independent diagonal form 
\be\dot{\chi }={\rm B} \chi,\label{Eq:eomchiLinear}\ee
whose solutions are the Floquet modes,
\be \chi _{\nu } \left(t\right)=\chi _{\nu } \left(0\right)e^{i\beta _{\nu } t}. \ee

We note that in the general case, $\Gamma$ mixes the coordinates $\vec{u}$ and their derivatives, and in addition is not unitary, although certain highly symmetric cases, e.g., as discussed in Secs.~\ref{Sec:PseudoLin} and \ref{Sec:PointsLin}, when the Mathieu equations decouple completely, are an exception.

\subsection{Solution Using an Expansion in Infinite Continued Matrix Inversions}\label{Sec:MathieuSolution}

We expand the solutions of the e.o.m \eqref{Eq:eomuInHomogenous} by using an analytical expansion which utilizes infinite-continued matrix inversions \cite{Risken}, see \cite{rfmodes} for details and further references. This expansion allows to obtain the frequencies and the coefficients of the solution vectors, to arbitrary precision. The solution is given in a form which is immediately suitable for obtaining the Floquet-Lyapunov transformation, as will be shown in the next subsection. 

We seek for solutions of eq.~\eqref{Eq:eomuInHomogenous}, in the form of a sum of two linearly independent complex columns vectors,
\be \vec{u} =\sum _{n=-\infty }^{n=\infty }\vec{C}_{2n} \left[b e^{i\left(2n+\beta \right)t} +c e^{-i\left(2n+\beta \right)t} \right] ,\label{Eq:uAnsatz}\ee 
where $b$ and $c$ are complex constants determined by the initial conditions. Stable modes will be described by $\beta$ taking a real nonintegral value (we exclude the case of integral $\beta$). Following the discussion in \textsection 4.70 of \cite{McLachlan}, for trapping parameters in the first stability zone of the Mathieu equation, $\beta$ can be chosen in the range $0<\beta<1$ for all stable modes.

By defining $R_{2n} =A-\left(2n+\beta \right)^{2}$ we can write infinite recursion relations for $\vec{C}_{2n}$,
\be Q\vec{C}_{2n-2} =R_{2n} \vec{C}_{2n} - Q\vec{C}_{2n+2}, \label{Eq:Recursion1}\ee 
and obtain two independent expansions in infinite continued matrix inversions
\be \vec{C}_{2} =T_{2,\beta }Q \vec{C}_{0} \equiv \left(\left[ R_{2} -Q\left[ R_{4} - Q\left[ R_{6} -... \right]^{-1}Q \right]^{-1}Q \right]^{-1}\right)Q \vec{C}_{0}. \label{Eq:MatrixInversions1} \ee
and
\be Q\vec{C}_{2} = R_{0} \vec{C}_{0} -Q\vec{C}_{-2} =\tilde{T}_{0,\beta }\vec{C}_{0} \equiv  \left(R_{0} -Q\left[ R_{-2} -Q\left[ R_{-4} -... \right]^{-1}Q \right]^{-1}Q \right)\vec{C}_{0}. \label{Eq:MatrixInversions2} \ee
Multiplying eq.~\eqref{Eq:MatrixInversions1} by $Q$ and defining
\be Y_{2,\beta } \equiv \tilde{T}_{0,\beta } - Q T_{2,\beta } Q\label{Eq:Y2beta}, \ee
we find that all characteristic exponents $\beta $ are zeros of the determinant of $Y_{2,\beta } $ (which is a function of $\beta $). If there are degenerate $\beta $'s they will appear as degenerate zeros of this determinant. The vector $\vec{C}_{0} $ for each $\beta $ is an eigenvector of $Y_{2,\beta } $ with eigenvalue $0$. Since $A$ and $Q$ are symmetric, $Y_{2,\beta } $ is symmetric as well, and so its kernel will be of dimension equal to the algebraic multiplicity of the $\beta$ root. The vector $\vec{C}_{2} $ can be obtained by an application of $T_{2,\beta }Q $ to $\vec{C}_{0} $, for $n=-1$ we use $\vec{C}_{-2} = \left[T_{-2,\beta }\right]^{-1}Q \vec{C}_{0}$, and so on for the other vectors. We note that the different vectors $\vec{C}_{2n,\beta } $ are not orthogonal in general, and the vectors at every order in $n$ mix different coordinates.

The general term of the expansion vanishes. Either $A$ or $Q$ may be singular and the expansion can still be applied in general. Even if both are singular, the expansion is valid if there are no integral values of $\beta$, a case which we do not tackle as noted above. Excluding perhaps isolated values of $\beta$ (and atypically in the $a-q$ parameter-space), all matrices which are inverted in the above expressions will be invertible, and while employing the algorithm in practice, the invertibility of the matrices is of course easily verified at each step. In Sec.~\ref{Sec:Numerics} we use in fact a generalization of the above expansion \cite{rfmodes} which includes also the next Fourier harmonic ($\cos 4t$, ommited from eq.~\eqref{Eq:KijFourier}).

\subsection{The Floquet-Lyapunov Transformation For Stable Modes}\label{Sec:FLTrans}

We now further assume that all Floquet modes are stable, i.e. that the $2f$ linearly-independent solutions of eq.~\eqref{Eq:eomFloquet} are oscillatory and thus come in complex conjugate pairs. This simplifies many expressions. We therefore take ${\rm B}$ of eq.~\eqref{Eq:B} in the block form
\be {\rm B}= \left(\begin{array}{cc} {iB} & { 0} \\ {0 } & {-iB }\end{array}\right), \qquad \left(B\right)_{f\times f} = {\rm diag}\left\{\beta_1,...,\beta_f\right\} \label{Eq:Bblock}\ee
where $\beta_j$ are positive. We define the $f$-dimensional matrix $U$ whose columns are constructed from the series of $f$-dimensional vectors $\vec{C}_{2n,\beta _{j}}$ obtained from the recursion relations for the solutions of eq.~\eqref{Eq:uAnsatz}, i.e.
\be \left(U\right)_{f\times f} = \left(\begin{array}{cc} 
{\sum \vec{C}_{2n,{\beta _{j} }} e^{i2nt} } & { ... } \end{array}\right)\ee
where in the above expression and for the rest of this section, the summation is over $n \in {\mathbb Z}$. We similarly define the $f$-dimensional matrix $V$ composed from column-vectors as
\be \left(V\right)_{f\times f} = \left(\begin{array}{cc} 
{i\sum \left(2n+\beta _{j} \right)\vec{C}_{2n,\beta _{j}} e^{i2nt}} & { ... } \end{array}\right). \ee

The matrices $U$ and $V$ can be chosen to obey the normalization condition
\be V^t\left(0\right)U\left(0\right) = \frac{1}{2}i,\label{Eq:Normalization}\ee
which is a rescaling imposed by multiplication with a (diagonal) matrix, such that
\be U\to U\left(-2iV^t\left(0\right)U\left(0\right)\right)^{-\frac{1}{2}},\label{Eq:ScalingMatrix}\ee
and $V$ accordingly.
As shown in \cite{rfmodes}, the Floquet-Lyapunov transformation and its inverse can then be obtained in closed form, and are given block-wise by (where $U^*$ denotes the complex conjugate of the matrix $U$),
\be\Gamma \left(t\right)=\left(\begin{array}{cc} {U} & { U^*} \\ {V } & {V^* }\end{array}\right),\quad \Gamma^{-1} \left(t\right)=\left(\begin{array}{cc} {iV^\dag } & {-iU^\dag} \\ {-iV^t} & {iU^t}\end{array}\right).\label{Eq:Gamma}\ee

This transformation is a canonical transformation from the Hamiltonian coordinates $\vec{u}$ and their conjugate momenta $\vec{p}=\dot{\vec{u}}$, to the variables given by
\[ \chi = \left(\begin{array}{c} {\xi} \\ {\zeta} \end{array}\right),\]
where $\xi$ is the new $f$-dimensional vector of coordinates and $-i\zeta$ are the new conjugate momenta (we here break the notation a little). Using the realness of $\vec{u}$ and $\vec{p}$ it is easy to verify that $\xi = \zeta^*$. The time dependence of these modes is
\be \xi _{j} \left(t\right)=\xi _{j } \left(0\right)e^{i\beta _{j} t}, \qquad \zeta _{j} \left(t\right)=\zeta _{j } \left(0\right)e^{-i\beta _{j} t}. \ee


\section{Computation of the Modes of a `Peculiar' Crystal}\label{Sec:Numerics}

\begin{figure}[ht]
\center{ \includegraphics[width=\linewidth]{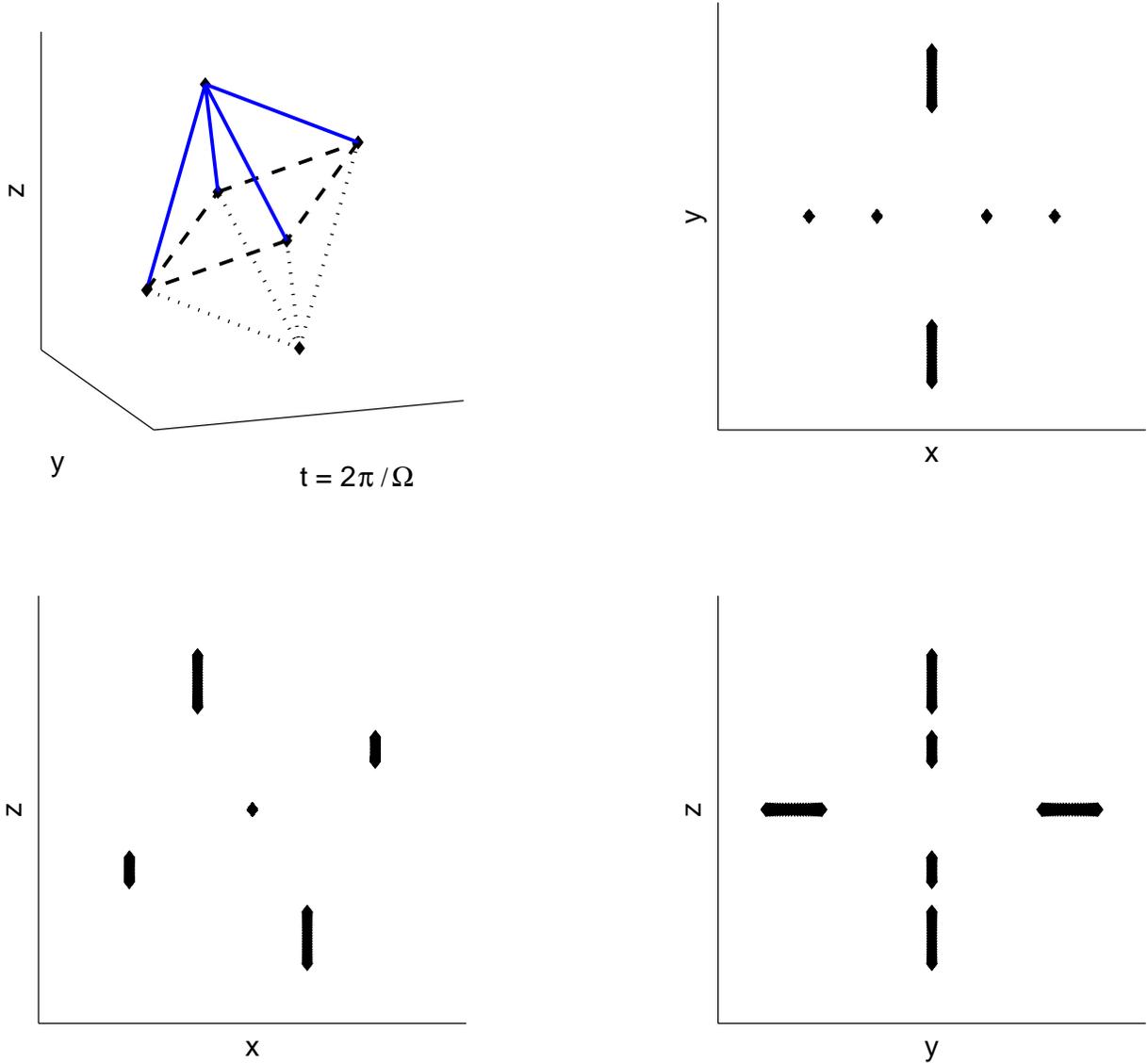}
\caption {The periodic `peculiar' crystal solution of 6 ions in a linear Paul at nearly-spherical trapping parameters (see text for details). The average locations correspond to a (nearly-regular) Octahedron. Upper-left : the ions at zero rf-phase. The other 3 figures show the ion trajectories over one rf-period. The oscillation with a large amplitude is given by eq.~\eqref{Eq:B2qB0} with an accuracy of $0.5\%$. The axial micromotion amplitude is of order $10^{-4}$ of the ions' radial positions (see Sec.~\ref{Sec:Periodic}).}
\label{Fig:Octahedron2}}
\end{figure}

\begin{figure}[ht]
\center{ \includegraphics[width=\linewidth]{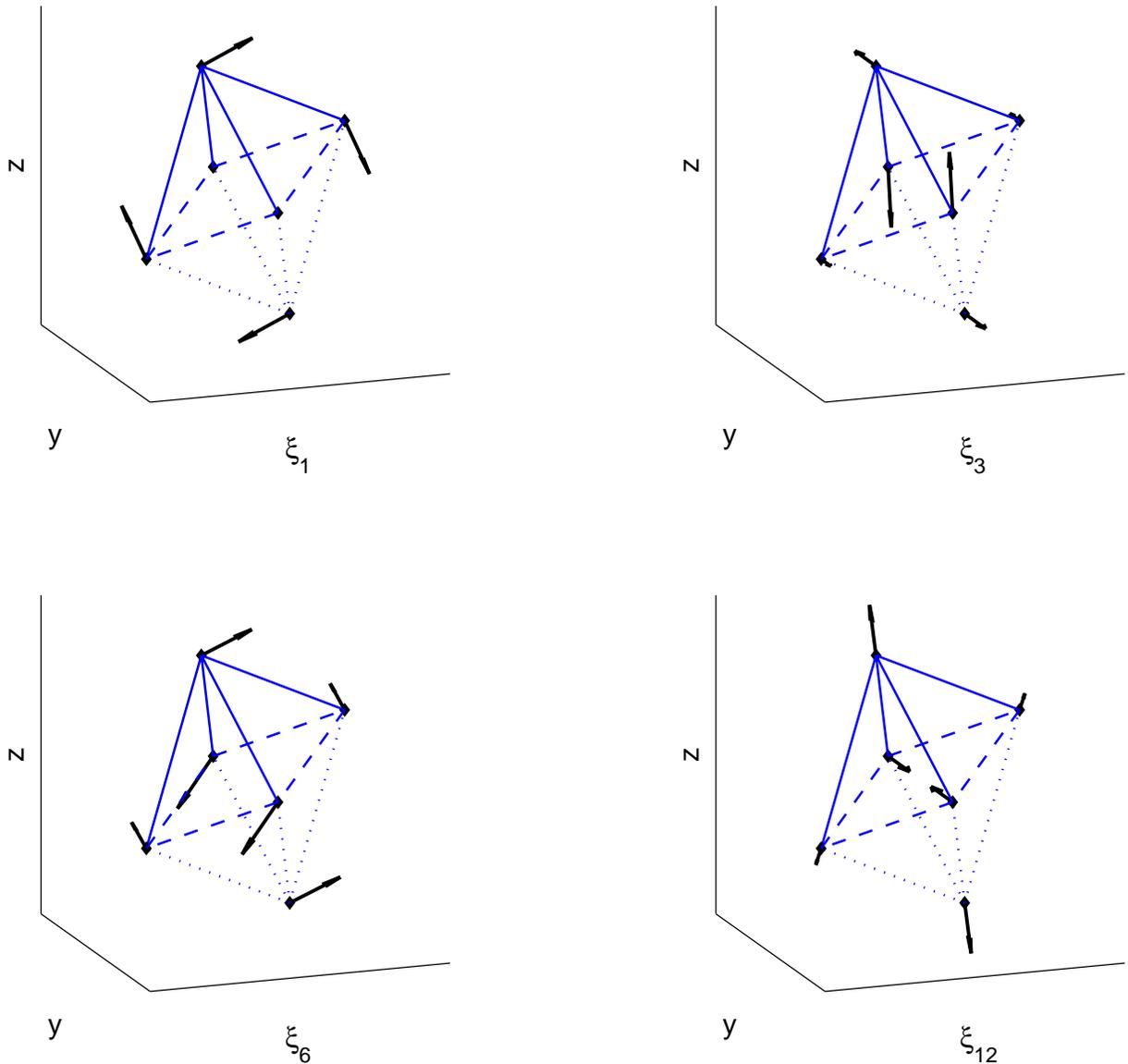}
\caption {Ions' primary direction of oscillation (given by $\vec{C}_0$), in four Floquet-Lyapunov modes, which are numbered from the lowest frequency, with $\xi_{1}$ being the lowest-frequency mode.}
\label{Fig:FloquetModeVectors}}
\end{figure}

\begin{figure}[ht]
\center{ \includegraphics[width=\linewidth]{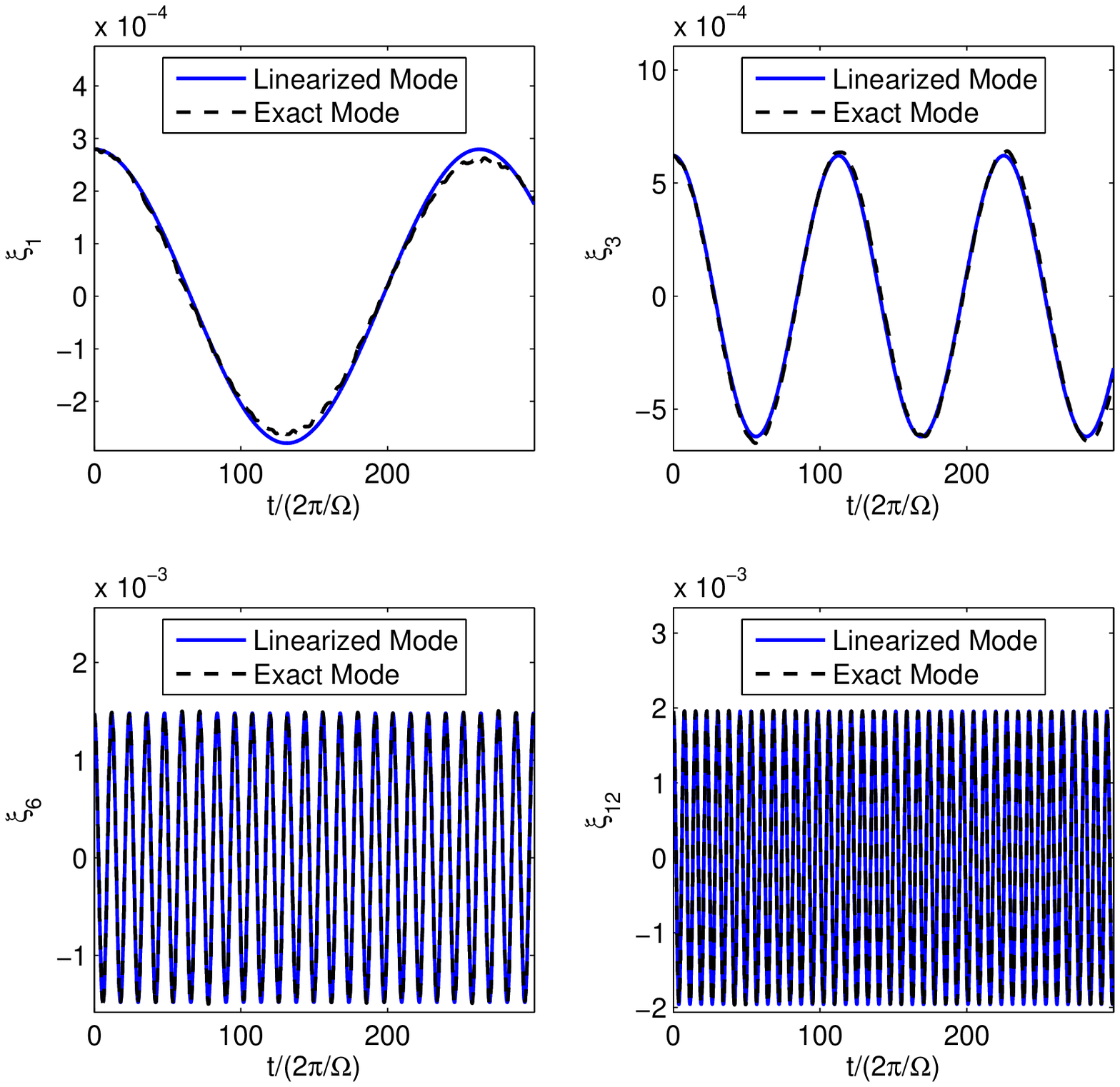}
\caption {Comparison between the analytical expansion and the exact numerical solution, for the Floquet-Lyapunov modes of Fig.~\ref{Fig:FloquetModeVectors}, with small random initial conditions. The modes are numbered from the lowest frequency, with $\xi_{1}$ being the lowest-frequency mode. Time is measured in periods of the rf frequency, and corresponding natural nondimensional units are used for distances.}
\label{Fig:FloquetMode300p}}
\end{figure}

In Figs.~\ref{Fig:Octahedron2}-\ref{Fig:FloquetMode300p} we consider a crystal of 6 ions, demonstrating the utility and generality of the analysis presented in this contribution. The simulation is of an almost ideal linear Paul trap (with only a 1\% DC asymmetry in the radial plane), such that the center-of-mass frequencies are degenerate to within 1\%. The Mathieu parameters are $q_y=0.41$ and $a_x=0.05766$. In the pseudopotential approximation, the corresponding minimum-configuration is a (nearly-regular) octahedron, with two ions sitting on each axis of the trap. The rf crystal on the other hand, may well deserve to be called `peculiar' (in the sense described in Sec.~\ref{Sec:PaulOverview}), since it is not oriented with the axes. There are two ions lying along the $y$-axis, along which the confinement is strongest, but the two other pairs of ions sit along lines which are rotated in $x-z$ plane. The large amplitude oscillation at the rf-period, and the absence of micromotion along the axial direction, confirm very accurately the results presented in Sec.~\ref{Sec:Periodic}.

Numerically, the periodic crystal solution can be obtained by starting from a simulation of the full e.o.m's with a friction (cooling) term and slowly turning it off (the adiabatic shutting-down of the damping term is important). The crystal is then followed for a period and the periodic solution and force matrix can be Fourier expanded to obtain the matrices $A$ and $Q$. Since the crystal is `peculiar', it has no corresponding pseudopotential limit (normal modes of regular polyhderons were investigated in \cite{PeetersSmallCoulombCrystals}, and see also references in \cite{Complex_plasmas}). For a very accurate description of the modes in the Paul trap, we add to eq.~\eqref{Eq:eomuInHomogenous} the next Fourier harmonic to get $\ddot{\vec{u}} +\left[A -2Q_2\cos 2t-2Q_4\cos 4t\right]\vec{u} = 0$, and expand it in the method of continued matrix inversions, using the formulas given in App.~B of \cite{rfmodes}, which generalize the formulas presented in Sec.~\ref{Sec:Solution}. This modification is required in order to obtain accurately the low frequency modes of nearly-degenerate configurations, as in a peculiar crystal.

\section{Concluding Comments}\label{Sec:Conclusion}

In this contribution we have investigated the dynamics of ion crystals in rf traps. We repeat the main results presented herein. In eq.~\eqref{Eq:rfeomModes} we show that in general, the rf couples the pseudopotential normal modes of the crystal, and also acts as a driving force. For a crystal configuration which has the same symmetry as the trapping potential (e.g. a chain of ions or a planar crystal in a trap with cylindrical symmetry), we find that the equations of motion for the modes become decoupled Mathieu equations as in eq.~\eqref{Eq:eomModes}, for which the stability analysis is trivial (but may be different from the pseudopotential approximation).

In eqs.~\eqref{Eq:B2qB0} and \eqref{Eq:B2q2B0} we derive results which have been observed in experiments and numerical simulations, namely that in a general ion crystal the micromotion amplitude in each coordinate is $q_{\alpha}/2$ of the respective average position, and that in the linear Paul trap, axial micromotion is negligibly small.

When the crystal solution differs from the pseudopotential limit (as in the peculiar crystals discussed above), or when a very accurate expansion of the modes is desired, the derivation of eq.~\eqref{Eq:eomuInHomogenous} takes into account the full rf crystal solution and ion interactions along the periodic trajectory. Sec.~\ref{Sec:Solution} describes briefly how to solve for the decoupled modes of oscillations of the ions, allowing obtaining explicitly the mode frequencies and solution vectors.

We have focused on single-species crystals in quadrupole traps, specifically the linear and hyperbolic Paul traps. However, generalization to other cases is easy. Crystals in multipole traps \cite{PhysRevA.75.033409,PhysRevA.80.043405,MartinaRing} can be treated by expanding the motion around a suitable periodic solution $\left\{\vec{R}_i^{\pi}\right\}$ as in Sec.~\ref{Sec:Periodic}. Keeping only the leading terms will lead to the equations treated above. Segmented traps and trap arrays \cite{SurfaceTrapLucas,MuirTrapArray,SchmidtKalerSegmented,Three_dimensional_lattice_of_ion_traps} can be handled simply by changing the Mathieu parameters felt by each ion (assuming that the rf drive has identical frequency). Crystals of nonidentical ions and other types of driving can be treated similarly, and various transformations \cite{Yakubovich} can be used to handle more general linear system similar to eq.~\eqref{Eq:eomuInHomogenous}, such as systems with first-order derivatives (e.g. linear damping, gyroscopic forces or magnetic fields \cite{Baltrusch,Peeters1D}). 

The framework presented here for calculating the classical normal modes and their frequencies can find immediate application in most studies involving trapped ion Coulomb crystals, including studies of how ion Coulomb crystals differ from uniformly charged liquid models \cite{DubinColdFluidPRL}, and possibly even crystalline beams \cite{SchaetzCrystallineBeams,CrystallineBeamPhonons}. The analysis is based on linearization of the nonlinear solution about a periodic solution, and the nonlinear correction terms can be written as a series expansion in the Floquet modes about the crystal solution. In a quadrupole trap the nonlinear terms would come only from the Coulomb interaction, and in a multipole trap, there will be terms coming from the nonlinear trapping potential. Such an expansion may serve as a starting point for study of nonlinear collective phenomena, e.g. the important phenomenon of rf heating (e.g. \cite{SchifferDrewsenHeating,HeatingSchuessler,HeatingSchiller,HeatingHasegawa} and many more).

In addition, an exact quantum description of the modes, and wavefunctions in the configuration space of the ions, are presented in \cite{rfmodes}, utilizing the Floquet-Lyapunov transformation described in Sec.~\ref{Sec:Solution}. The nontrivial time-dependent wavefunctions could, e.g., eventually become important for understanding trapped ion chemistry at ultracold temperatures \cite{ultracold_molecules}.
\section*{Acknowledgments}

BR acknowledges the support of the Israel Science Foundation and the European Commission (PICC). BR and AR acknowledge the support of the the German-Israeli Foundation. MD acknowledges financial support from the Carlsberg Foundation and the EU via the FP7 projects `Physics of Ion Coulomb Crystals' (PICC) and `Circuit and Cavity Quantum Electrodynamics' (CCQED). HL wishes to thank R. Geffen and M. Brownnutt.

\bibliographystyle{../hunsrt}

\bibliography{../bibfile}

\end{document}